\documentclass[preprint,12pt]{elsarticle}

\usepackage{amssymb}
\usepackage{amsmath}
\usepackage{graphicx}

\journal{Optics Communications}

\begin{document}

\begin{frontmatter}



\title{Adiabatic Transparency of Multilevel Atomic Media
 for Short High-intensity Pulses}


\author[1,2]{G. Grigoryan\fnref{a}}
\ead{gaygrig@gmail.com}
\author[1]{V. Chaltykyan}
\author[1]{E. Gazazyan}
\author[2]{A. Hovhannisyan}
\author[2]{O.Tikhova}

\address[1]{Institute for Physical Research, 0203 Ashtarak-2, Armenia}
\address[2]{Russian-Armenian (Slavonic) University}

\begin{abstract}
We consider a medium of multilevel atomic systems interacting with radiation pulses. A relatively simple technique of analytic calculations is proposed, which allows revealing all necessary conditions (with sufficient conditions to be checked separately) imposed on the interaction parameters, for which the mean dipole moment of a multilevel atomic medium vanishes, i.e., the medium becomes transparent via adiabatic interaction. The proposed technique is based on the method of quasienergies and illustrated for three- and five-level atomic systems. The necessary conditions for the propagation length where the interaction adiabaticity is preserved in the medium are obtained.
\end{abstract}
\end{frontmatter}

\section{Introduction}
Microscopic theory of coherent resonant interaction between radiation and individual atoms is sufficiently complete \cite{1,2,3} but in case of macroscopic volumes of interaction a number of questions arises which are up today open. First, this is the question of spatial scale where the processes occurring with individual atoms can be realized. Then, this is the question of the energy exchange between a pulse and the medium and between different pulses travelling in the medium simultaneously. Joint action of nonlinear and dispersive properties of a resonant medium leads to modification of temporal and spectral characteristics of an optical pulse and to loss of optical information contained in the incident pulses.
Adiabatic processes constitute a specific sub-class of coherent interactions, which rely on adiabatic evolution of the quantum system \cite{4,5}. As the evolution is slow, the system typically remains in a certain laser-controlled quantum state, which is an eigenstate of the interaction Hamiltonian. This yields a relatively stable situation. However, even at adiabatic interaction the nonlinear dispersion of the medium leads in general to parametric broadening of spectra of pulses \cite{6}, which in its turn results in rather fast destruction of the interaction adiabaticity \cite{7} . The possibility of compensation for the medium dispersion was revealed experimentally as early as in 1973 in case of medium of three-level V-type atoms \cite{8}. After that the possibility of soliton regime of pulse propagation in such media was studied in detail \cite{9}. This mechanism of compensation for the nonlinear dispersion of medium is based on destructive interference of dipole moments induced by radiation at the adjacent transitions of medium atoms.
Another striking example of bleaching of resonant medium via adiabatic interaction is EIT \cite{10,11}  which is based on the so-called dark state formed in a three-level system \cite{12}. The dark state does not involve the intermediate state because of quantum interference; due to this fact the field-induced dipole moments vanish leading to medium bleaching.
The purpose of the present work was to find the necessary (but not always sufficient) conditions for similar bleaching of a medium consisting of atomic systems with more complicated level diagrams. We describe a relatively simple technique of analytic calculation of adiabatic propagation of light pulses and illustrate this technique for a five-level atomic system. One of the above-mentioned mechanisms (or their combination) underlies the adiabatic bleaching to be considered. We also obtain the necessary conditions for the propagation length where the interaction adiabaticity does not break in the medium.
\label{}
\section{Formalism}
We assume that considered pulses have durations much shorter than all times of relaxation and at the same time much longer than the inverse frequency distance between the closest quasienergies (eigenvalues of the interaction Hamiltonian), in order to ensure the adiabaticity of interaction.
Adiabatic propagation of pulses in an atomic medium is described by a self-consistent system of truncated equations in running coordinates $x$ and $t-x/c$; these equations may be represented in the form \cite{13,14}
\begin{equation}\label{1}
    \frac{\partial E_j}{\partial x} = -i \frac{2 \pi N \omega_j \hbar}{c} \frac{\partial \lambda}{\partial E^{*}_j}
\end{equation}
Here $\lambda$ is the eigenvalue of the interaction Hamiltonian, $E_j$ the complex amplitude of the pulse field, and other notations are conventional. If the system has a quasienergy which remains constant during the overall time of interaction, i.e., $\partial \lambda/\partial E^*_j=0$, then according to \eqref{1}, the induced dipole moment in such a system is zero and the pulse propagates in such medium without change in shape at the group velocity equal to $c$. It is known \cite{15} that the quasienergies (eigenvalues) of the system are the roots of characteristic equation
\begin{align}\label{2}
    det(H-\lambda I)=0
\end{align}
where $I$ is the unit matrix and $H$ the interaction Hamiltonian (in frequency units) which may in the resonance approximation be represented in the form
\begin{equation}\label{3}
    H=\sum \sigma_{i,i}\delta_{i-1}-(\sum \sigma_{i,i+1} \Omega_i+h.c.)
\end{equation}
with $\sigma_{i,j}$ being the projection matrices, $\Omega_i$ the Rabi frequencies of corresponding transitions $i\longrightarrow i+1$, and $\delta_{i-1,1}$ the ($i$-1)-photon detuning ($\delta_0$=0) (Definition of multi-photon detunings depends on the specific scheme of interaction, see Appendix). The Rabi frequencies are assumed to be real and positive while their phases, which can vary when travelling in the medium, are included into one-photon detunings ($ \Delta_{i}=\omega_{i+1,i}-\omega_{i}+\dot{\varphi_{i}}$, if $\omega_{i+1,i}>0$ and $\Delta_i=\omega_{i,i+1}-\omega_{i}+\dot{\varphi_{i}}$, if $\omega_{i+1,i}<0 $). By differentiating the characteristic equation with respect to $E^{*}_{j}$ and setting to zero the derivatives of quasienergy, we obtain a system of algebraic equations, from which the needed values of interaction parameters may be determined.
The problem can, however, be simplified. Really, at turning off the fields the interaction Hamiltonian is diagonalized and the roots of equation \eqref{2} go to the following constant values:
\begin{equation}\label{5}
    \lambda_i\longrightarrow\delta_{i-1}
\end{equation}

The problem of determination  of necessary (but not always sufficient) conditions for medium transparency is thus reduced to the search of such parameters of adiabatic interaction for which the system quasienergy remains equal to one of resonance detunings always during interaction. By substituting these values successively into equation \eqref{2} we obtain the needed conditions for the interaction parameters.
\begin{figure}
  \includegraphics[width=10cm]{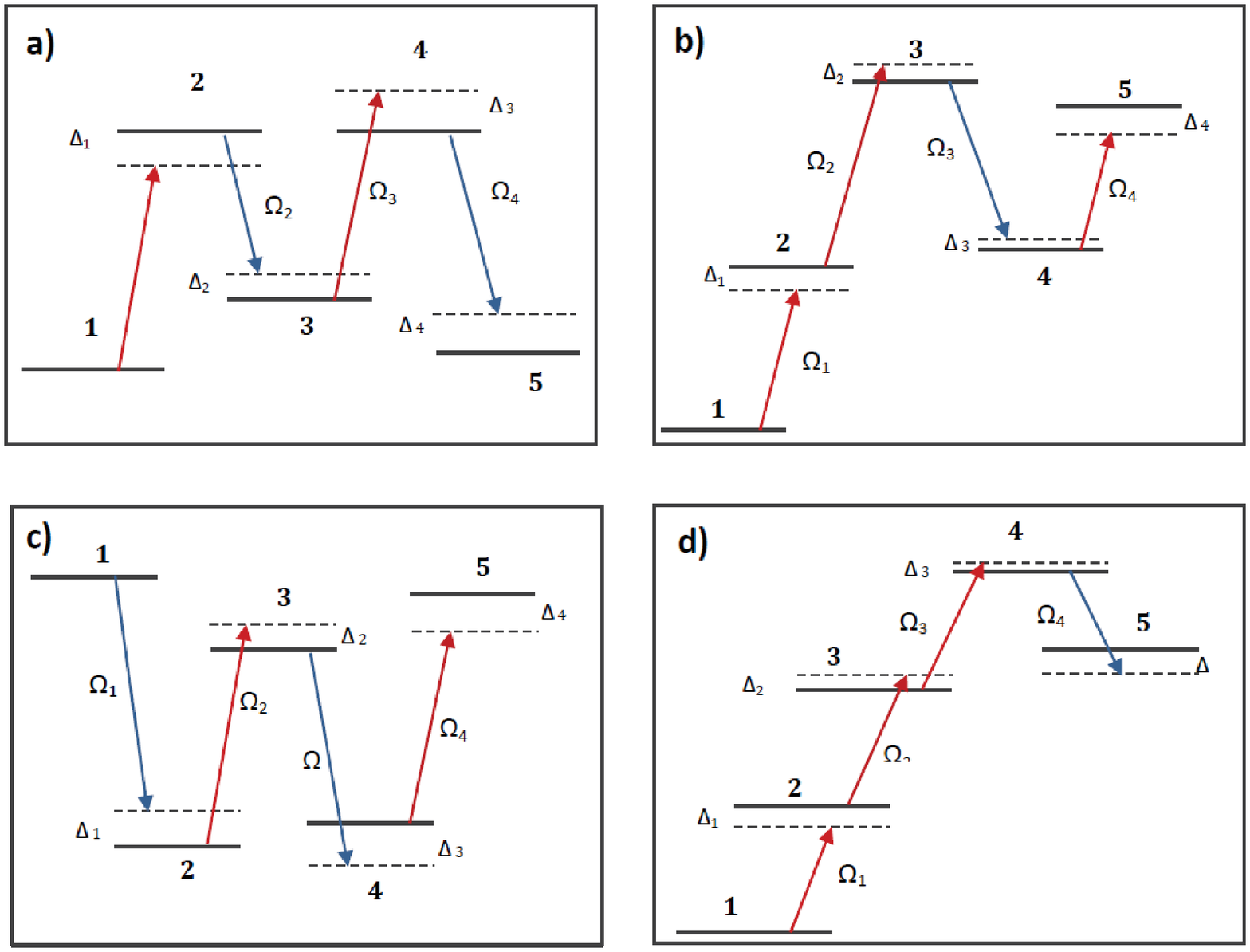}\\
  \caption{Several five-level configurations and relevant Rabi frequencies. }\label{1}
\end{figure}

For example, in case of a three-level system the root $\lambda=0$ is realized at exact two-photon resonance $\delta_2=0$ and under the same condition the root $\lambda=\delta_2=0$ is realized. The wave function corresponding to these quasienergies (dark state) is well known \cite{11}:
\begin{equation}\label{6}
    |\psi\rangle=\cos\theta |1\rangle-\sin\theta|3\rangle
\end{equation}
where $\tan\theta=\Omega_1/\Omega_2$. As the population of the level $|2\rangle$ is zero, the dipole moments of $1\longrightarrow2$ and $2\longrightarrow3$ transitions also vanish.
The third root $\lambda=\delta_1$ is realized at degenerate pumping ($\Omega_1=\Omega_2$) and the condition for detunings $\delta_2=2\delta_1$. As distinct from the previous case, the wave function involves all bare atomic states:
\begin{equation}\label{7}
    |\psi\rangle=\frac{1}{\sqrt{2}}\sin\Phi|1\rangle+\cos\Phi|2\rangle-\frac{1}{\sqrt{2}}\sin\Phi|3\rangle
\end{equation}
where $\tan\Phi=\sqrt{2}\Omega_1/\delta_1$.The medium becomes transparent because of destructive interference of dipole moments of transitions $1\longrightarrow2$ and $2\longrightarrow3$. It should be noted that in case of non-degenerate pumping the root $\lambda=\delta_1$ does not provide transparency, since the dipole moments of the transitions $1\longrightarrow2$  and $2\longrightarrow3$ do not equal zero and they do not interfere. So, the condition $\lambda=\delta_1$ is in this case necessary, but not sufficient.

Now consider a five-level atomic system and the pulses resonant with only the adjacent transitions. Several examples of such diagrams are demonstrated in Fig.1. We show in what follows that in such systems six regimes of adiabatic bleaching of medium are possible.

a)Under conditions of exact two-photon resonances($\delta_2$=0, $\delta_4=0$) a dark state is realized in the system (see, e.g., \cite{16}) which corresponds to the root $\lambda$=0:
\begin{equation}\label{8}
     |\psi\rangle=\frac{1}{R}(\cos\theta_1\cos\theta_2|1\rangle-\sin\theta_1\cos\theta_2|3\rangle+\sin\theta_1\sin\theta_2|5\rangle)
\end{equation}
where $\tan\theta_1=\Omega_1/\Omega_2; \tan\theta_2=\Omega_3/\Omega_4, R=\sqrt{(1-\sin^2\theta_2\cos^2\theta_1)}$ , This superposition does not involve the levels 2 and 4 (destructive interference) and the system is well studied in literature for the case $\Omega_1=\Omega_3$ , $\Omega_2=\Omega_4$ \cite{16,17} (see Fig.2).

b) The value $\lambda=\delta_1$ can be realized under conditions $\delta_1=\delta_4=\delta_2/2$ and degenerate pumping at the transition $1\longrightarrow2\longrightarrow3$  ($\Omega_1=\Omega_2=\Omega$). The wave function does not contain the level 4 and the dipole moments of transitions $1\longrightarrow2$ and $2\longrightarrow3$ are in antiphase leading to their destructive interference:
\begin{equation}\label{9}
         |\psi\rangle=\frac{1}{R}(\sin\Phi\cos\theta_2|1\rangle+\sqrt{2}\cos\Phi\cos\theta_2|2\rangle-\sin\Phi\cos\theta_2|3\rangle+\sin\Phi\sin\theta_2|5\rangle)
\end{equation}
where $R=\sqrt{\sin^{2}\Phi\sin^{2}\theta_{2}+2\cos^{2}\theta_{2}}$ and $\tan\Phi=\sqrt{2}\Omega/\delta_{1},\tan\theta_{2}=\Omega_{3}/\Omega_{4}$.
This wave function passes to \eqref{7} at $\theta_2 =0$. For realization of state \eqref{9} the atom should initially be prepared in state $|2\rangle$. Besides, the pulses at transitions $3\longrightarrow4$ and $4\longrightarrow5$ must be turned on in counterintuitive sequence, i.e., $\theta_2 (t\longrightarrow-\infty)\longrightarrow0$. If also $\theta_2$ ($t\longrightarrow\infty)\longrightarrow 0$, the atom returns after interaction into state $|2\rangle$.But if $\Omega_4$ is turned off prior to $\Omega_3$ ($\theta_2 (t\longrightarrow\infty)\longrightarrow \pi/2$), the atom passes into state  $|5\rangle$. So, with use of counterintuitive sequence of pulses at transitions  $3\longrightarrow4$ and $4\longrightarrow5$ with a longer pulse at transition $1\longrightarrow2\longrightarrow3$ (see Fig.3), it is possible to transfer the population from level 2 to level 5. In this case a population inversion is produced between the levels 5 and 4, which may be used for amplification of radiation. This state may efficiently be used in atomic level configurations of Fig. 1c (W-system). With the same conditions the value $\Delta=\delta_4$ may be realized.

c) The value $\lambda=\delta_1$, discussed above, may also be realized under different conditions, e.g., at  $\delta_1=\delta_2=\delta_3, \delta_4=2\delta_1$ and degenerate pumping of transitions $1\longrightarrow2\longrightarrow3$ and $ 3\longrightarrow4\longrightarrow5$ ($\Omega_1=\Omega_2$ and  $\Omega_3=\Omega_4$). (Note that the same conditions lead to realization of the values  $\lambda=\delta_2$ and  $\lambda=\delta_3$). In this case the wave function involves all bare atomic states, but dipole moments of both degenerate transitions are zero(as can easily be seen from Eqs \eqref{10} below):
\begin{align*}
  |\psi\rangle=\frac{1}{R}(\sin\Phi_1\sin\Phi_2|1\rangle+\cos\Phi_1\sin\Phi_2|2\rangle-\sin\Phi_1\sin\Phi_2|3\rangle-
\end{align*}
\begin{align}\label{10}
     -\sin\Phi_1\cos\Phi_2|4\rangle+\sin\Phi_1\sin\Phi_2|5\rangle)
\end{align}
where $R=\sqrt{\sin^{2}\Phi_{1}+\sin^{2}\Phi_{2}+\sin^{2}\Phi_{1}\sin^{2}\Phi_{2}}$ and $\tan\Phi_1=\Omega_1/\delta_1,\tan\Phi_2=\Omega_3/\delta_3$.The atom should be initially in state $|2\rangle$  and the pulses turn on in counterintuitive sequence. At the end of interaction the atom turns to be in state $|4\rangle$. This state may efficiently be used in atomic level configuration in Fig. 1d for exitation of the hight-lying level. The same state will  be obtained when  $\Omega_1=\Omega_4$ and  $\Omega_3=\Omega_2$.

d) The value  $\lambda=0$ also may be realized in different ways. For example, instead of  $\delta_2=\delta_4=0$ we can require vanishing of three-photon detuning,  $\delta_3=0$, and  $\delta_2=\delta_4\neq0$. In this case, if pumping of the $3\longrightarrow4\longrightarrow5$ transition is degenerate ($ \Omega_3=\Omega_4$), we obtain
\begin{equation}\label{11}
    |\Psi\rangle=\frac{1}{R}(\sin\Phi_3\cos\theta_1|1\rangle-\sin\Phi_3\sin\theta_1|3\rangle+\cos\Psi_3\sin\theta_1|4\rangle+\sin\Psi_3\sin\theta_1|5\rangle)
\end{equation}
where $R=\sqrt{\cos^2\Phi_3+\sin^2\theta_1}$ and $\tan\Phi_3=\Omega_4/\delta_4$, $\tan\theta_1=\Omega_1/\Omega_2$ . As distinct from the dark state \eqref{8}, it contains four out of five atomic states. For realization of this state the atom should be prepared in state $|4\rangle$.

e,f) Similarly, the value  $\lambda=\delta_2$ can be obtained under conditions $2\delta_2=\delta_1+\delta_3=\delta_4, \Omega_1=\Omega_4, \Omega_3=\Omega_2$ and the value  $\lambda=\delta_3$ under conditions  $\delta_1=\delta_2, 3\delta_3=\delta_2+\delta_4, \Omega_1=\Omega_2.$
\begin{figure}
  \includegraphics[width=15cm]{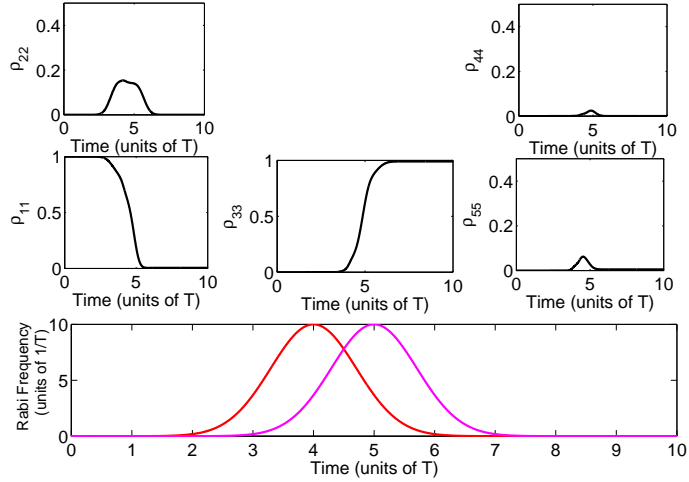}\\
  \caption{Dynamics of atomic level populations ($\rho_{jj}$) in M-system and the sequence of pulses. Shapes of pulses are chosen to be Gaussian and the value of the one-photon detuning is $\Delta$=10/T.}\label{123}
\end{figure}

\section{Propagation}
Although in adiabatic approximation all the above-considered schemes lead, as follows from Eq.\eqref{1}, to adiabatic bleaching of the medium, we cannot determine the criterion of interaction adiabaticity for a single atom knowing only a single value of quasienergy. Since we are interested in the propagation problem, we must determine the adiabaticity criterion for the medium and it may be found directly from the propagation equations. By combining the Schr\"{o}dinger equation with truncated Maxwell equations, we obtain for a medium consisting of M-type atoms (Fig.1a) the following equations describing variations of one-photon detunings (frequencies of pulses) and pulse intensities (in running coordinates $x$, $\tau=t-x/c$) during propagation in medium:
\begin{align*}
  \frac{\partial\Omega^2_1}{\partial x} =q_1\frac{\partial|b_1|^2}{\partial\tau}
\end{align*}
\begin{align*}
    \frac{\partial\Omega^2_2}{\partial x} =-q_2\frac{\partial(|b_1|^2+|b_2|^2)}{\partial\tau}
\end{align*}
\begin{equation}\label{12}
     \frac {\partial\Omega^2_3}{\partial x} =q_3\frac{\partial(|b_1|^2+|b_2|^2+|b_3|^2)}{\partial\tau}
\end{equation}
\begin{align*}
    \frac{\partial\Omega^2_4}{\partial x} =q_5\frac{\partial|b_5|^2}{\partial\tau}
\end{align*}
\begin{align*}
    \frac{\partial\Delta_i}{\partial x} =q_i\frac{\partial}{\partial\tau}\frac{Re(b^*_i b_{i+1})}{\Omega_i}
\end{align*}
where $q_i=2 \pi N \omega_i |d_{i,i+1}|^2/\hbar c $ ($d_{i,i+1}$ is the matrix element of the dipole moment of respective transition).
For other atomic level schemes equations for intensities differ in only the signs of derivatives with respect to populations. For example, in case of ladder (cascade) configuration we have to change the signs in the second and fourth equations for field intensities \eqref{12}. The choice of sign is determined by whether the field is absorbed or emitted in the specific transition. If the pumping is degenerate (i.e., if the pulses at adjacent transitions have the same frequency), the right-hand-sides of corresponding equations should be added.
Since it is the time derivatives that stand in rhs of these equations, by use for atomic populations the expressions following from the wave functions obtained above, we take into account the first nonadiabatic correction which lead to the change in spectral and temporal characteristics of pulses. Correspondingly, the conditions of smallness of these changes are criteria of adiabaticity in the medium. We will now demonstrate this statement for a traditional STIRAP-chain in a M-system and for a W-system (whose dynamics of travelling in the medium is not studied to the best of our knowledge).
\section{Traditional STIRAP-STIRAP-chain in M-system.}
We will restrict ourselves to the case of equal strengths of adjacent transitions and to the conditions $\Omega_1=\Omega_3$ and $\Omega_2 =\Omega_4$. With use of wave function \eqref{8} we obtain directly from Eqs.\eqref{11}
\begin{align*}
    \frac{\partial \Omega^2_1}{\partial x}=q \frac{\partial}{\partial\tau} (|b_1|^2-|b_5|^2)
\end{align*}
\begin{equation}\label{13}
    \frac{\partial \Omega^2_2}{\partial x}=q \frac{\partial}{\partial\tau} (|b_5|^2-|b_1|^2)
\end{equation}
\begin{align*}
    \frac{\partial\Delta_i}{\partial x}=0
\end{align*}

As follows from the last equation, the self-phase modulation in the medium is absent, i.e., the conditions for multiphoton detunings given at the medium entrance are preserved during propagation. The quantity $\Omega^2_1+\Omega^2_2=\Omega^2_0$ is the integral of motion. Hence, the entire dynamics of propagation of pulses in medium is described by the angle $\theta(x,t)$. Since $\Omega^2_1=\Omega^2_0\sin^2\theta$ and $\Omega^2_2= \Omega^2_0\cos^2\theta$, we finally obtain for the mixing angle  $\theta(x,t)$ the equation
\begin{equation}\label{14}
    \frac{\partial\theta}{\partial x}+\frac{qf(\theta)}{\Omega^{2}_{0}}\frac{d\theta}{\tau}=0
\end{equation}
where $f(\theta)=(1+2\cos^{2}\theta\sin^{2}\theta)/(1-\cos^{2}\theta\sin^{2}\theta)^2$.
As follows from the obtained equation, the variation of the angle $\theta(x,t)$ can be neglected at the lengths of propagation satisfying the condition
\begin{equation}\label{15}
    \frac{qx}{\Omega^2_0 T} f(\theta)<\frac{3qx}{\Omega^2_0 T}<<1
\end{equation}
This is just the necessary criterion of the adiabaticity of interaction in the medium, so that at the travelling lengths satisfying the condition \eqref{15} the influence of the medium on the process of coherent population transfer may be disregarded.
Equation \eqref{14} may be solved analytically also without this limitation. The solution determined by, e.g., the method of characteristics may be represented as
\begin{equation}\label{16}
    \theta(x,t)=\theta_0(\tau-x/u)
\end{equation}
where $\theta_0$ is a function given at the medium entrance and $u$ the nonlinear group velocity defined by the equation
\begin{equation}\label{17}
    \int^{\tau}_{\tau-x/u}\Omega^2_0d\acute{t}=qxf(\theta)
\end{equation}
\begin{figure}
  \includegraphics[width=16cm]{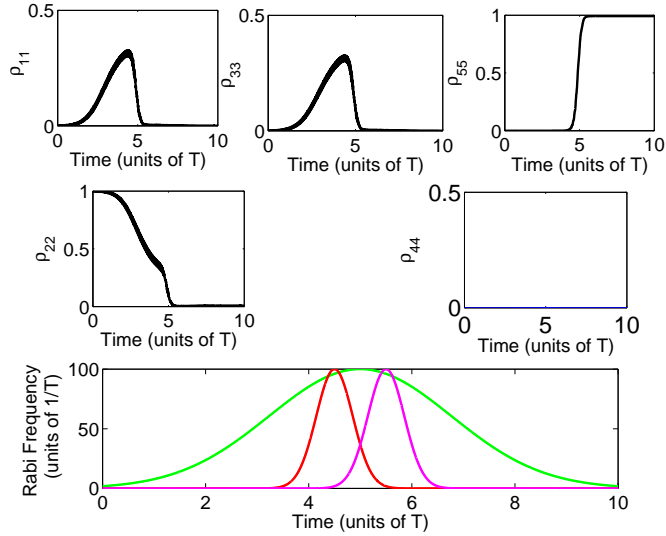}\\
  \caption{Dynamics of atomic level populations ($\rho_{jj}$) in W-system and the sequence of pulses. Shapes of pulses are chosen to be Gaussian and the value of the one-photon detuning is  $\Delta$=10/T.}\label{3}
\end{figure}
Because of the nonlinear group velocity the time derivative of the function $\theta$ will increase infinitely at the travelling lengths $qxf(\theta)/\Omega^2_0T\sim1$ which leads at this lengths to breaking of interaction adiabaticity. Note that in the regime linear with respect to probe field (small mixing angles) the function $f(\theta)$ goes to unity and the shock wave does not form which allows using this system for information storage.
\section{Population transfer in W-system}
For a W-system in case where the wave function \eqref{8} is realized, we obtain for intensities and frequencies of pulses the following system of equations (transition strengths are assumed, as above, to be equal):
\begin{align*}
    \frac{\partial\Omega^2_1}{\partial x} = q \frac{d}{d\tau} |b_2|^2
\end{align*}
\begin{equation}\label{18}
    \frac{\partial\Omega^2_3}{\partial x} = - \frac{\partial\Omega^2_4}{\partial x} =q \frac{d}{d\tau} |b_5|^2
\end{equation}
\begin{align*}
    \frac{\partial\Delta_i}{\partial x} =0
\end{align*}
The self-phase modulation is, as in previous case, absent, since in transition $2\longrightarrow1,3$ (V-system) conditions of compensation for dispersion are met and in transition $3\longrightarrow4\longrightarrow5$ (inverted $\Lambda$ -system) the conditions for EIT are satisfied. The quantity $\Omega^2_3+\Omega^2_4=\Omega^2_0$  is the integral of motion. Using the definition of mixing angles $\theta_2$, we obtain $Ô$ directly from \eqref{18} the following equation:
\begin{equation}\label{19}
    \frac{\partial\Omega^2_1}{\partial x}+\frac{1}{u_1} \frac{\partial\Omega_1}{\partial t} =\Omega_1 A\frac{ \partial\tan^2\theta_2}{\partial t}
\end{equation}
where $u_1$ is the nonlinear group velocity, always lower than $c$, which is defined as
\begin{equation}\label{20}
    \frac{1}{u_1}=\frac{1}{c}+ \frac{2q\delta^2_1(2+\tan^2\theta_2)}{(\delta^2_1+\Omega^2_1(2+\tan^2\theta_2))^2}
\end{equation}
The rhs of Eq.\eqref{19} is caused by depletion of the field $\Omega_1$ via transformation of photons at  $\Omega_1$ into photons of the field $\Omega_3$. The factor $A$ in \eqref{19} equals
\begin{equation}\label{21}
    A=\frac{2q\delta^2_1}{(\delta^2_1+\Omega^2_1(2+\tan^2\theta_2))^2}
\end{equation}
We see that at the lengths in the medium satisfying the inequality
\begin{equation}\label{22}
    qxT/(\delta_{1}T)^2<<1
\end{equation}
either of these effects is negligibly weak. Before turning on the field $\Omega_3$ (i.e., at $\theta_2=0$, see Fig.3) the solution of the Eq. \eqref{18} is
\begin{equation}\label{23}
    \Omega^{2}_{1}(t,x) = \Omega^{2}_{01}(t-x/u_1)
\end{equation}
Inequality \eqref{22} has in this case a simple physical meaning: the group delay of the pulse in the medium is much shorter than its duration. After turning off the field $ \Omega_4$ (i.e., at $\theta_2=\pi/2  $) we again obtain the same solution. At the lengths $4qx\Omega^2_1/\delta^2_1T_1\sim1$  the nonlinear group velocity will lead to formation of the shock-wave front, i.e., to breaking of interaction adiabaticity ($T_1$ is duration of $\Omega_4$ pulse).The full dynamics of propagation of the field $\Omega_1$ requires joint solution of the equation \eqref{19} and the equation for the mixing angle $\theta_2$:
\begin{equation}\label{24}
    \frac{\partial\theta_2}{\partial x}-\frac{4q}{\Omega^2_1} f_1(\theta_2,\Phi) \frac{d\theta_2}{d\tau}=\frac{qB}{\Omega^2_0}\frac{\partial\Omega^2_1}{\partial\tau}
\end{equation}
where
\begin{align*}
    f_1(\theta_2,\Phi)=\frac{\cos\theta_2\sin^2\Phi}{(2cos^2\theta_2+\sin^2\theta_2\sin^2\Phi)^2}
\end{align*}
\begin{equation}\label{25}
    B=\frac{sin2\theta_2\cos^2\Phi}{(2\cos^2\theta_2+\sin^2\theta_2\sin^2\Phi)^2}\frac{1}{(\delta^{2}_{10}+2\Omega^2_1)}
\end{equation}
The travelling length where the variation of the angle $\theta_2$ may be neglected should satisfy the condition
\begin{equation}\label{26}
    qxT_1/(\Omega_0T_1)^2<<1
\end{equation}
Denominators of expressions \eqref{22} and (25) provide criteria of interaction adiabaticity for a single atom, while the whole expressions give criteria of adiabaticity for the medium, i.e., they determine the travelling lengths where the propagation effects may be neglected. The medium  parameter $qxt$ may be expressed in terms of center of line absorption coefficient $\alpha_0=q/\Gamma , qxT_1=\alpha_0 x\Gamma T_1.$
Then, under condition $\Gamma <<\delta_{1}$ the expression \eqref{22} may be rewritten as
\begin{align*}
        \frac{\alpha_{0} \Gamma^2 x}{\delta^2_{10}}<<\Gamma T<<1
\end{align*}
which means that the linear absorption in medium, caused by the atomic linewidth, is negligibly small. However, by integrating the first equations \eqref{17} over time we obtain for the pulse energy $W_i=\int^\infty_\infty \Omega^2_idt:$
\begin{align*}
    \frac{dW_1}{dx}=q_1(|b_2|^2(+\infty)-|b_2|^2(-\infty))
\end{align*}
In case of complete transfer of population we have $|b_2|^2(+\infty)-|b_2|^2(-\infty)=-1$. Hence,$W_1 (x)=-q_1 x+W_0$, with $W_0$  being the pulse energy at the medium entrance. So, the length at which $q_1 x=W_0$ is the length of full depletion (absorption) of pumping radiation .
So, at the propagation length satisfying the condition
\begin{align*}
    \frac{qx}{\Omega^2_{01}T_1}<\frac{qx}{\delta^2_{1}T_1}<<1
\end{align*}
we obtain full absorption of the field $\Omega_1$ caused by the presence of the field $\Omega_2$.

The main distinction of the equation \eqref{24} describing the dynamics of propagation of the mixing angle $\theta_2$ from the similar equation \eqref{14} for the M-system, is the fact that the nonlinear velocity of propagation of angle $\theta_2$ exceeds the speed of light, so the population transfer in this system is superluminal \cite{18}. This situation is similar to that of superluminal population transfer in a tree-level system via bright state considered in \cite{18}. Really, by neglecting the rhs of Eq.\eqref{23} in case of large one-photon detuning, we obtain a solution similar to that obtained above for the M-system, but with replacing x by  -x:
\begin{equation}\label{26}
    \theta(x,t)=\theta_0(\tau+x/v),
    \int^{\tau}_{\tau+x/v}\Omega_0dt=-2qxf_1(\theta_2,\Phi)
\end{equation}
More detailed analysis of propagation of pulses with taking into account the first nonadiabatic correction requires combined solution of the system of equations \eqref{19} and \eqref{24}; it is out of scope of the present work.
\section{Acknowledgment}
Work was in part supported by the Ministry of Education and Science of Armenia (MESA), grant no. 11-1c124 of State Science Committee of MESA, IRMAS International Associated Laboratory,ANSEF-optPS-2911 and the Volkswagen Stiftung I/84 953.
\section{Conclusion}
We have developed and analyzed, for the example of a five-level system, a relatively simple method for finding necessary conditions under which adiabatic short light pulses can travel in the medium without distortion of the shape and phase. It was shown that the necessary conditions for such bleaching of medium (i.e., vanishing of dipole moments induced in the medium at the frequencies of all interacting fields) is the equality of system quasienergies to one of the values of single- or multi-photon detunings of resonances. In this case the medium bleaching may be caused by both interference of quantum states and interference of dipole moments of different transitions if pumping is degenerate. The existence of such regimes means that at the propagation lengths where the interaction adiabaticity does not break, stable superposition states are produced all over the medium and these states may easily be controlled by adjusting the parameters of pulses. Finding of the sufficient conditions requires determination of roots of algebraic equations which may be done only numerically for the order of equation higher than three. However, the sufficiency of found necessary conditions may easily be checked directly by calculation of the induced dipole moments.
In addition, we demonstrated for the examples of M- and W-systems the possibility to find a generalized criterion of adiabaticity of interaction in the medium without determination of all eigenvalues of interaction Hamiltonian; we analyzed population transfer in these systems and its peculiarities during propagation in the medium.
\bibliographystyle{elsarticle-num}
.
\end{document}